\documentclass[aps,pra,showpacs,amssymb,superscriptaddress,nofootinbib,twocolumn,notitlepage,10pt]{revtex4-1}

\usepackage{amsmath,amsfonts,amsthm,amssymb}
\usepackage{setspace}
\usepackage{amsmath}
\usepackage{appendix}
\usepackage[ansinew]{inputenc}
\usepackage{bbm}
\usepackage{bm}
\usepackage{amsbsy}
\usepackage{dsfont} 
\usepackage{graphicx} 
\usepackage{epsfig}
\usepackage{epstopdf}
\usepackage{dsfont}
\usepackage{color}
\usepackage[colorlinks]{hyperref}
\usepackage[figure,table]{hypcap}
\usepackage{enumerate}
\usepackage[resetlabels]{multibib}
\usepackage{notes2bib}

\hypersetup{
	bookmarksnumbered,
	pdfstartview={FitH},
	citecolor={darkgreen},
	linkcolor={darkred},
	urlcolor={darkblue},
	pdfpagemode={UseOutlines}}
\definecolor{darkgreen}{RGB}{50,190,50}
\definecolor{darkblue}{RGB}{0,0,190}
\definecolor{darkred}{RGB}{238,0,0}
\usepackage{soul}

\renewcommand{\d}{\dagger}
\newcommand{\Tr}{\mathrm{Tr}}

\newcommand{\ii}{\ensuremath{\mathrm{i}}}
\newcommand{\ee}{\ensuremath{\mathrm{e}}}
\newcommand{\bra}[1]{\ensuremath{\left\langle#1\right\rvert}}
\newcommand{\ket}[1]{\ensuremath{\left\lvert#1\right\rangle}}

\newcommand{\dd}{\mathrm{d}}

\renewcommand{\Re}{\mathrm{Re}}
\renewcommand{\Im}{\mathrm{Im}}

\begin{document}

\title{Coherence-assisted single-shot cooling by quantum absorption refrigerators}
\author{Mark T. Mitchison}
\email{marktmitchison@gmail.com}
\affiliation{Quantum Optics and Laser Science Group, Blackett Laboratory, Imperial College London, London SW7 2BW, United Kingdom}
\affiliation{Clarendon Laboratory, University of Oxford, Parks Road, Oxford OX1 3PU, United Kingdom}
\author{Mischa P. Woods}
\email{mischa.woods@googlemail.com}
\affiliation{Centre for Quantum Technologies, National University of Singapore, 3 Science Drive 2, Singapore 117543}
\affiliation{University College of London, Department of Physics \& Astronomy, London WC1E 6BT, United Kingdom}
\author{Javier Prior}
\email{javier.prior@upct.es}
\affiliation{Universidad Polit\'{e}cnica de Cartagena, C/Dr Fleming S/N 30202 Cartagena, Spain}
\author{Marcus Huber}
\email{entangledanarchist@gmail.com}
\affiliation{Departament de F\'{i}sica, Universitat Aut\`{o}noma
de Barcelona, E-08193 Bellaterra, Spain}
\affiliation{ICFO-Institut de Ciencies Fotoniques, Mediterranean
Technology Park, 08860 Castelldefels (Barcelona), Spain}
\begin{abstract}
The extension of thermodynamics into the quantum regime has received much attention in recent years. A primary objective of current research is to find thermodynamic tasks which can be enhanced by quantum mechanical effects. With this goal in mind, we explore the finite-time dynamics of absorption refrigerators composed of three quantum bits (qubits). The aim of this finite-time cooling is to reach low temperatures as fast as possible and subsequently extract the cold particle to exploit it for information processing purposes. We show that the coherent oscillations inherent to quantum dynamics can be harnessed to reach temperatures that are colder than the steady state in orders of magnitude less time, thereby providing a fast source of low-entropy qubits. This effect demonstrates that quantum thermal machines can surpass classical ones, reminiscent of quantum advantages in other fields, and is applicable to a broad range of technologically important scenarios.
\end{abstract}
\maketitle

\section{Introduction}

The development of classical thermodynamics in the 19th century underpinned the Industrial Revolution, and the enormous economic growth and social changes that followed. Now, in the 21st century, the burgeoning quantum technological revolution promises unprecedented advances in our computation and communication capabilities, enabled by harnessing quantum coherence. As our machines are scaled down into the quantum regime, it is of prime importance to understand how quantum mechanics affects the operation of these devices. This problem has attracted great interest to the field of quantum thermodynamics over the last few years. 

One useful approach in this regard is to explore simple physical models which highlight novel aspects of quantum thermal machines. The quantum absorption refrigerator is the quantum extension of a classical machine devised in the 19th century (see, for example, Ref.~\cite{Levy2012prl} and references therein). The smallest possible model with couplings between physical particles and thermal reservoirs was first studied by Linden et al.~\cite{Linden2010prl}, who considered a three-qubit refrigerator. While the history of studying these machines dates back a long time even in the quantum regime, this three-qubit model was the first where the role of quantum information resources was studied~\cite{Brunner2013}, revealing that entanglement in the steady state prohibits achieving perfect Carnot efficiency, but potentially increases cooling efficiency. A wide variety of other quantum absorption refrigerator models have also been proposed in the recent literature~\cite{Levy2012pre,Correa2014pre,Gelbwaser2014pre,Leggio2015pra,Silva2015pre}.

Designing thermodynamic processes that can be enhanced by quantum dynamics is a pivotal challenge in the field of quantum thermodynamics. One of the paradigmatic thermodynamic tasks concerns work efficiency at the quantum scale. Here already the very definition of quantum mechanical work is debated~\cite{Allahverdyan2008,Horodecki2013a,Skrzypczyk2014}, yet in different scenarios quantum mechanical advantages seem possible~\cite{DeLiberato2011,Perarnau2015prx, Brandner2015njp, Uzdin2015prx}. Another avenue in this endeavour is the engineering of the environment itself to enhance quantum processes~\cite{Correa2014,Ross2014,Killoran2015jcp,Lim2015natcomm}. We circumvent the potential controversies regarding the practical value of work and efficiency by concentrating on a different figure of merit, namely the achievable temperature, and by using thermal baths to drive the refrigerator, therefore needing no notion of work.

Previous research on the quantum aspects of heat engines and refrigerators has focused almost exclusively on their operation in the steady state.  However, in many applications one wishes to reach low temperatures as rapidly as possible, in which case understanding the short-time behaviour is essential. In particular, the cooling may be applied only transiently, after which the cold object is extracted for use. As a somewhat frivolous yet illustrative example, consider the problem of refrigerating a beverage on a hot day. Maximum enjoyment is obtained if the beverage can be cooled and then consumed quickly, at a temperature significantly lower than that of the environment. A more serious example could be the initialisation of a register of qubits for quantum information processing, where the aim is to produce states with high purity (low entropy). Fast cooling is advantageous here since it may reduce the overall time taken to complete the quantum information protocol. Both of these situations exemplify what we call \textit{single-shot cooling}: the one-time application of a refrigeration device in order to considerably and rapidly reduce the temperature of the object in question.

In the present work, we study the application of three-qubit absorption refrigerators to single-shot cooling. By considering variations of the basic processes underlying energy dissipation and transport, we elucidate the role of coherence in the operation of such a device. Furthermore, we demonstrate that dramatic improvements can be obtained in both the cooling time and the achievable temperatures by taking advantage of coherent oscillations that appear in the transient dynamics of the refrigerator. Most importantly, the presence of coherence in the device allows one to reach lower temperatures than the steady state. This use of quantum coherence in order to cool below the steady-state temperature we refer to as \textit{quantum single-shot cooling}.

Note that several physical implementations of quantum thermal machines have recently been proposed~\cite{Abah2012prl,Chen2012epl,Mari2012prl,Venturelli2013prl,Zhang2014prl,Dechant2015prl}, and experimental efforts to construct such devices in the laboratory are currently under way. Our findings could be relevant for future experiments, which may be practically limited by finite coherence times. From a more fundamental perspective, our scheme provides one of the first examples in which quantum coherence plays an active and necessary role in improving the performance of machines driven only by thermal noise.

\section{Theoretical Framework}

\subsection{Description of the refrigerator}

We consider a quantum absorption refrigerator comprising three qubits described by standard Pauli operators $\sigma^{x,y,z}_i$, with the local Hamiltonian
\begin{equation}\label{qubitH}
H_{\mathrm{loc}} = \frac{1}{2} \sum_{i=1}^3 E_i\sigma^z_i.
\end{equation}
Throughout this article, we employ units of energy, time and temperature such that $E_1 = 1$, $\hbar = 1$ and $k_B = 1$. The qubits are coupled together according to a three-body interaction
\begin{equation}
\label{interactionH}
V = g \ket{010}\bra{101} + \mathrm{h.c.},
\end{equation}
where the computational basis states $\ket{0}$, $\ket{1}$ denote the eigenvectors of $\sigma^z$. In order for this interaction to be energy-conserving, in the sense that $[H_\mathrm{loc},V] = 0$, we demand that the qubit energies satisfy $E_2 = E_1 + E_3$. The interaction \eqref{interactionH} then drives resonant transitions within the \textit{transport subspace} spanned by the states $\ket{010}$ and $\ket{101}$. 

Each qubit $i$ is in contact with an independent heat bath $B_i$, which drives it towards a thermal equilibrium state at inverse temperature $\beta_i = 1/T_i$, where $T_1 \leq T_2 < T_3$. When the system parameters are judiciously chosen, the dynamics propels qubit 1 (the ``cold qubit") towards a new temperature $\tilde{T}_1 < T_1$. It is in this sense that the system behaves as a refrigerator, whose basic operating principle may be understood as follows. The interaction Hamiltonian \eqref{interactionH} couples the cold qubit to a ``virtual qubit"~\cite{Brunner2012pre} comprising the $\{\ket{01}$,$\ket{10}\}$ subspace of the Hilbert space of qubits 2 and 3. When these qubits are in equilibrium with their respective baths, the populations of the virtual qubit states are thermally distributed at an effective temperature given by
\begin{equation}
\label{effTemp}
T_v = \frac{E_2 - E_3}{\beta_2 E_2 -\beta_3 E_3}.
\end{equation}
If the parameters of the refrigerator are arranged so that $T_v < T_1$, then the cold qubit equilibrates to a lower temperature as it exchanges energy with the virtual qubit on the approach to the steady state.

In the following, we specialise to the case where $T_1 = T_2 =: T_r$, and we define $T_h := T_3$. This is in many ways the most natural scenario, where we have a single source of free energy --- namely the hot bath at temperature $T_h$ --- enabling us to cool below the ambient ``room temperature" $T_r$. So long as qubits 1 and 2 are sufficiently spatially separated~\bibnote{More precisely, the distance between the two qubits must be much greater than the correlation length $\xi = c/\Omega$, where $c$ is a characteristic velocity of bath excitations and $\Omega$ is the frequency cut-off characterising the bandwidth of environmental noise.}, we can treat the effect of their common environment as arising from two independent baths~\cite{Palma1996prs}.

\begin{figure}
\includegraphics[scale=0.3]{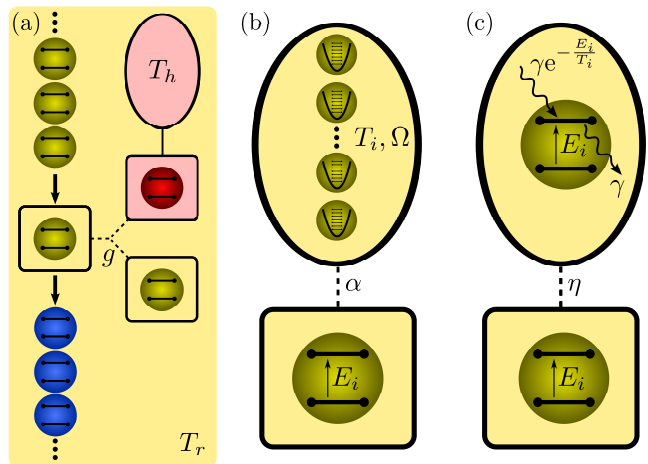}
\caption{Cartoon illustrating our theoretical set-up. (a) A stream of thermal qubits passes sequentially through a quantum absorption refrigerator. Each thermal qubit interacts with the refrigerator for a finite time, and exits the device at a lower temperature. (b) The canonical model of thermal reservoirs (Model I) consists of an infinite collection of harmonic oscillators, characterised by a bandwidth $\Omega$ and a dimensionless coupling strength $\alpha$. (c) An alternative bath model (Model II) consists of a single fictitious qubit undergoing Markovian gain and decay processes, characterised by a linewidth $\gamma$ and a dimensionless coupling strength $\eta$. \label{fig:cartoon}}
\end{figure}

The general set-up that we have in mind is illustrated in Fig.~\ref{fig:cartoon}(a). We assume that all three qubits are initially in thermal equilibrium with their respective reservoirs. The initial state is therefore 
\begin{equation}
 \rho_A(0) = \bigotimes_{i=1}^3 \frac{\ee^{-\beta_iE_i\sigma^z_i/2}}{\Tr\left (\ee^{-\beta_iE_i\sigma^z_i/2}\right )},
\end{equation} 
where $\rho_A(t)$ denotes the reduced state of the refrigerator qubits at time $t$. At $t = 0$ the interaction \eqref{interactionH} between the qubits is switched on. The aim is then to \textit{prepare} the cold qubit in a low-entropy state. (We emphasise that this is quite different from the usual objective considered by previous authors, namely to \textit{maintain} the cold qubit at a low temperature.) Once the lowest achievable temperature has been reached at time $t_0$, our cooling protocol is complete, and the cold qubit may be extracted for use, for example, in quantum information processing. Classically, we usually expect that the lowest achievable temperature occurs only as $t\to \infty$. The existence of a finite optimal extraction time $t_0$ is the essence of the quantum single-shot cooling protocol. If we suppose now that there exists a large supply of qubits thermalised at the ambient temperature $T_r$, we can perform the same procedure repeatedly to produce a steady stream of low-entropy qubits. 

\subsection{Models of thermalisation}

In order to quantitatively analyse the refrigerator we must specify a thermalisation model. On the other hand, one would like to obtain general results that are independent of any particular model. In order to avoid being too restricted by our assumptions, we employ two different approaches to modelling the thermal baths. In each case, we assume that the baths are Markovian (memory-less) and weakly coupled to the refrigerator, and we describe the dynamics by a master equation in Lindblad form. In the following paragraphs, we describe these models in a qualitative way, deferring the full details to Appendix~\ref{appendixA}. 

\subsubsection{Model I}

The canonical procedure to model Markovian reservoirs is to suppose that each qubit is coupled to an infinite collection of harmonic oscillators spanning a broad range of frequencies (Fig.~\ref{fig:cartoon}(b)). In our case, the baths are described by identical Ohmic spectral functions of the form
\begin{equation}
\label{spectralFunction}
J(\omega) = \alpha \omega \ee^{-\omega/\Omega}.
\end{equation}
This function quantifies the strength of the coupling between each qubit and the oscillators near frequency $\omega$, weighted by the density of states of the reservoir (see Appendix~\ref{appendixA}). The effect of the baths is therefore characterised by two parameters: a dimensionless coupling strength $\alpha$, and a frequency cut-off $\Omega$ leading to a bath memory time of order $\Omega^{-1}$. Markovian dynamics is obtained when $\Omega$ is much larger than all other frequency scales and $\alpha \ll 1$, so that the dissipation rates are much smaller than the natural frequencies $\{E_i\}$ of the qubits.

Some care must be taken when treating the effect of the inter-qubit coupling on the thermalisation dynamics. We are able to derive two different master equations depending on the magnitude of $g$. The first equation is valid in the strong-coupling limit, where $g$ is much larger than the dissipation rates~\cite{Correa2013pre}. The second master equation holds in the weak-coupling limit, when $g$ is comparable to or smaller than the dissipation rates~\cite{Wichterich2007pre,Rivas2010njp}. 

\subsubsection{Model II}

An alternative bath model consists of assigning to each qubit of the refrigerator an additional fictitious qubit, which is damped by a perfectly Markovian thermal bath at temperature $T_i$, such that the spontaneous emission rate is $\gamma$ (Fig.~\ref{fig:cartoon}(c)). Excitations are then allowed to hop between each fictitious qubit and its associated physical qubit at a rate $\eta\gamma$. This simulates an effective thermal reservoir with a memory time of order $\gamma^{-1}$, coupled with strength $\eta$ to the physical qubit. Markovian dynamics of the physical qubits are therefore obtained when $\eta \ll 1$ and $\gamma$ is larger than all other frequency scales. The energy splitting of each fictitious qubit is chosen to be resonant with the frequency $E_i$ of the physical qubit, which ensures that each physical qubit in isolation is driven towards a local thermal state at temperature $T_i$. 

This model enables us to explore the strong-coupling limit of large $g$ without invoking the rotating-wave approximation, which involves a time-averaging over the autonomous dynamics of the refrigerator. This time-averaging has been shown to lead to unphysical predictions in certain scenarios where the open system is coupled to multiple heat baths at different temperatures~\cite{Wichterich2007pre}. Thus Model II provides us with an independent check on the validity of our results.

\section{Results}

\subsection{Short-time dynamics of the refrigerator}

We now present our quantitative results, obtained by numerical solution of the equations of motion. Our first observation is that sufficiently strong coherent coupling between the qubits drives damped Rabi oscillations of the local qubit populations. However, unlike Rabi oscillations due to local driving fields, the three-body interaction \eqref{interactionH} does not induce any local coherences between the qubit populations. The reduced state of each qubit is diagonal at all times and may therefore be assigned an effective temperature 
\begin{equation}
\label{effTemp}
\tilde{T}_{i}(t) = \frac{-E_i}{2\tanh^{-1}\langle\sigma^z_i(t)\rangle}.
\end{equation}

\begin{figure}
\includegraphics[scale=0.25]{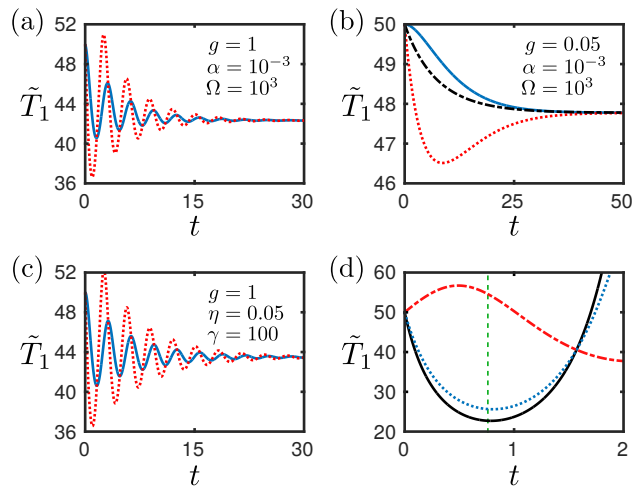}
\caption{Effective temperature dynamics of the cold qubit, with $E_1 = 1$, $E_2 = 2$, $T_r = T_1 = T_2 = 50$ and $T_h = T_3 = 100$. (a-c)~Blue solid lines indicate the evolution of an initial thermal product state, red dotted lines depict the evolution with 1\% of the maximum coherence added to the initial state. (a)~Model I in the strong-coupling regime. (b)~Model I in the weak-coupling regime, plotted with the corresponding stochastic model (black dot-dashed line). (c) Model II in the strong-coupling regime. (d)~Short-time effective temperature dynamics of an ensemble of cold qubits, where coherence with some phase noise is added to the initial state. The magnitude of the initial coherence is 5\% of the maximum and the phase uncertainties are $\delta\phi = 0$ (black solid line), $\delta\phi = \pi/4$ (blue dotted line) and $\delta\phi = \pi$ (red dot-dashed line). The temperature minimum in the absence of phase noise is shown by the green vertical dashed line. The data shown in (d) are calculated using Model I with the same parameters as (a). \label{Figure2}}
\end{figure}

To illustrate this feature of the short-time dynamics, we plot several examples of the evolution of the cold qubit temperature in Fig.~\ref{Figure2}, which demonstrate that the Rabi oscillations allow the cold qubit to reach lower temperatures than the steady state. In the strong-coupling regime we find damped temperature oscillations with the approximate period $\pi/g$. Therefore, the optimal quantum single-shot cooling procedure consists of extracting the cold qubit after a time $t_0 \approx \pi/(2g)$. So long as the coupling $g$ is larger than the relaxation rate, the same qualitative behaviour is found in both Model I (solid line in Fig.~\ref{Figure2}(a)) and Model II (solid line in Fig.~\ref{Figure2}(c)). In contrast, when $g$ is much smaller than the relaxation rate, these oscillations are over-damped so that no temperature minimum occurs in a finite time (solid line in Fig.~\ref{Figure2}(b)). Nevertheless, we have found that quantum single-shot cooling is possible over a very broad range of parameters, so long as the coupling $g$ is significantly larger than the relaxation rate.

The physical origin of the temperature oscillations is the exchange of energy between the refrigerator qubits due to the interaction \eqref{interactionH}. Our second important observation is that this energy transport is driven by coherence in the transport subspace. This can be seen straightforwardly by examining the Heisenberg equations of motion for the local energy expectation values $h_i = E_i \langle \sigma^z_i\rangle/2$. The resulting expression for the cold qubit is of the form
\begin{equation}
\label{energyBalance}
\frac{\dd h_1}{\dd t} = \dot{\mathcal{Q}}_1(t) - 2g E_1\Im \mathcal{C}(t).
\end{equation}
This equation represents an energy balance between the rate of heat absorbed from the bath $\dot{\mathcal{Q}}_1(t)$ and the coherent flow of energy into the other qubits, which is proportional to the imaginary part of the coherence in the transport subspace:
\begin{equation}
\label{coherence}
\mathcal{C}(t) = \Tr \left [\rho(t) | 010 \rangle\langle 101 |\right ],
\end{equation}
where $\rho(t)$ denotes the quantum state. Eqs.~\eqref{energyBalance} \& \eqref{coherence} give a direct link between the presence of coherence and the flow of energy across the refrigerator. We expect an analogous relationship to hold for any quantum refrigerator in which energy transport between distinct sub-systems occurs via coherent Hamiltonian evolution. 

Note that the same mechanism, whereby coherence enhances the flow of energy, has recently been shown to lead to an operational advantage for heat engines and power refrigerators working in the steady state \cite{Uzdin2015prx}. We also mention the similarity of our protocol with algorithmic cooling \cite{Boykin2002pnas,Baugh2005nature}. Indeed, the three-qubit unitary swap operation we employ is formally identical to that proposed for specific algorithmic cooling protocols~\cite{Remppthesis2007}, which may also be used in a single-shot operation similar to our proposal. The novel feature of our set-up is that no work is required to perform cooling. Rather, the free energy source is provided by the temperature difference between the hot thermal bath at temperature $T_h$ and the ambient environment at temperature $T_r$. This temperature difference establishes a population bias between the states $\ket{101}$ and $\ket{010}$, which enables the unitary generated by $V$ to redistribute excitations, thereby cooling the target qubit.

\subsection{Effect of initial coherence}

To further elucidate the fundamental role of coherence in energy transport, we consider a situation where some coherence in the transport subspace is added to the initial state, without modifying the thermally distributed populations. In order to ensure the positivity of the quantum state, the magnitude of this coherence is upper bounded by $|\mathcal{C}(0)|<	\mathcal{C}_\mathrm{max}$, where
\begin{equation}
\label{cohUpperBound}
\mathcal{C}_\mathrm{max} = \prod_{i=1}^3 \frac{1}{2} \mathrm{sech} \left (\frac{\beta_i E_i}{2}\right ).
\end{equation}
For each example in Figs.~\ref{Figure2}(a-c) we have also plotted the dynamics with a very small amount of initial coherence $\mathcal{C}(0) = \ii \mathcal{C}_\mathrm{max}/100$. We find that the amplitude of the temperature oscillations is noticeably enhanced in all three cases. In the weak-coupling example plotted (red dotted line in Fig.~\ref{Figure2}(b)), the initial coherence gives the most dramatic advantage, since now the temperature minimum occurs in finite time. Initial coherence is actually a necessary ingredient for quantum single-shot cooling in this case and thus should be considered as an additional non-classical resource to the free energy of the hot bath. 

This result reinforces the notion that coherence is a useful resource in quantum thermodynamics, which has lately been highlighted by a number of other authors. Indeed, it has recently been shown that coherence can improve the performance of thermal machines in the alternative context of a time-dependent Hamiltonian \cite{Uzdin2015_2}. In the abstract resource-theoretic approach to quantum thermodynamics, coherence is a resource which is non-increasing under any allowed operation \cite{Lostaglio2015,ciwilinski}, and from which work can be extracted \cite{Perarnau2015prx,Kamil2015}. Coherence may also play the role of a catalyst, but only for infinite-dimensional quantum systems \cite{Aberg2014}. Furthermore, it has been demonstrated explicitly that work can be extracted from coherence by quantum measurements \cite{Kammerlander2015}.

Note that, according to Eq.~\eqref{energyBalance}, the phase of $\mathcal{C}(0)$ determines the direction of the initial flow of energy into the qubit. Therefore, it is even possible, for example, to heat the qubit by adding initial coherence with the opposite phase, even when the system behaves as a refrigerator in the steady state. Likewise, one can transiently cool even when the steady-state behaviour is that of a heat pump. Similar phase effects have recently been described in Ref.~\cite{Oviedo2015}, where the authors show that bath fluctuations can revert the detailed balance condition in certain open quantum systems, creating a net flux of energy from the environment into the system.

Due to the aforementioned sensitivity of the dynamics to the phase of the coherence, it is important to consider the impact of any phase uncertainty resulting from imperfect initial state preparation. Specifically, we suppose that the initial coherence is given by $\mathcal{C}(0) = \ii r\ee^{\ii\phi}$, where $\phi$ is a zero-mean random variable.  Assuming that the cooling protocol is performed repeatedly on multiple qubits, the relevant quantity to consider is the temperature of the ensemble of cold qubits. For the rest of this section it should be understood that the term temperature refers to the property of the ensemble. (A characterisation of the temperature fluctuations of individual qubits is beyond the scope of this article.) 

The introduction of phase noise leads to two effects: an increase of the minimum temperature, and a shift in the time at which this minimum occurs. In Appendix~\ref{appendixB} we provide an approximate analytical quantification of the impact of phase noise, assuming that the phase fluctuations are small and approximately Gaussian. We have also numerically investigated the case of non-Gaussian phase noise, in particular the case where $\phi$ is uniformly distributed in the range $\phi \in [-\delta\phi/2,+\delta\phi/2]$. Some examples of the resulting temperature dynamics of the cold qubit ensemble are plotted in Fig.~\ref{Figure2}(d). 

We find that a narrow phase noise distribution leads to a small shift in the minimum temperature (e.g.\ blue dotted line in Fig.~\ref{Figure2}(d)).  The most deleterious impact is obtained when the phase noise distribution is very broad, so that the temperature minimum is shifted to a much later time. Extracting the cold qubit at the expected minimum may actually lead to heating of the cold qubit (red dot-dashed line in Fig.~\ref{Figure2}(d)). On the other hand, it may be possible to predict the dynamics of the ensemble temperature in advance, e.g.\ because the phase noise distribution is known. In this case it is possible to extract the cold qubit at the appropriate time to obtain some temperature advantage, unless this time is so late that thermal dissipation damps away the coherent oscillations.

\subsection{Stochastic absorption refrigerator}

In the previous two sections we presented strong evidence that coherence is a useful resource that may be harnessed to produce a significant advantage to single-shot cooling. On the other hand, we can show that coherence is not a necessary ingredient for quantum absorption refrigerators operating in the steady state. This is because energy transport can also be described by a stochastic process in which excitations are transferred incoherently between the qubits. Such a classical description is appropriate in the presence of strong dephasing~\cite{Kamiya2015ptep}, which in our case comes directly from the thermal baths (see Ref.~\cite{Uzdin2015prx} for an alternative dephasing model in the context of power refrigerators). 

As a concrete example, in Appendix~\ref{appendixC} we derive an effective master equation describing the asymptotic relaxation of the populations in the computational basis, valid when the coupling $g$ is much smaller than the dissipation rates. We show that energy transport in this limit can be effectively modelled by stochastic transitions between the states $\ket{101}$ and $\ket{010}$ at a rate $2g^2/\Gamma$, where $\Gamma$ is a characteristic dephasing rate due to the action of the thermal baths (see Eq.~\eqref{bigGamma}). 

The effective master equation accurately describes the long-time temperature dynamics of the cold qubit on its approach to the steady state (black dot-dashed line in Fig.~\ref{Figure2}(b)). This result demonstrates that three-qubit refrigerators are capable of cooling even when their dynamics is entirely classical. We see from Fig.~\ref{Figure2}(b) that there is no difference between the quantum and classical model in terms of the temperatures which can be achieved in the steady state. 

In the transient regime, on the other hand, the temperature dynamics are qualitatively different in the quantum and classical cases. Classical refrigeration models described by stochastic rate equations generically exhibit pure exponential relaxation, precluding the possibility of single-shot cooling below the steady-state temperature. This lies in stark contrast to the generic behaviour we have found in strongly coupled quantum absorption refrigerators, i.e.\ temperature oscillations enabling us to rapidly achieve temperatures lower than the steady state.

\subsection{Comparison with steady-state cooling}

\begin{figure}
\includegraphics[scale=0.29]{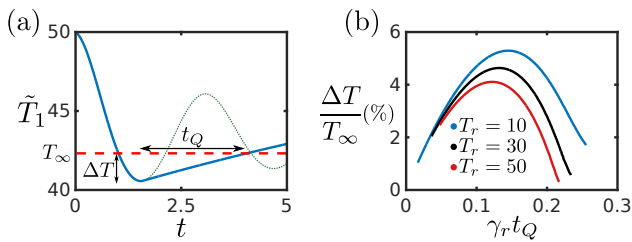}
\caption{(a) Example evolution using the same parameters as Fig.~\ref{Figure2}(a), with (blue solid line) and without (green dotted line) a switch-off of the interaction at time $t_0 = \pi/(2g)$, illustrating the quantum temperature and time advantages $\Delta T$ and $t_Q$, respectively. The steady-state temperature $T_\infty$ is shown by the red dashed line. (b) Trade-off between the fractional temperature advantage and the time advantage (in units of the relaxation time $\gamma_r^{-1}$), with the same parameters as (a) apart from the temperatures. Each line shows the results for a fixed room temperature $T_r$ and a range of hot bath temperatures $T_h \in [T_r+1, T_r+200]$. The value of $t_Q$ decreases as $T_h$ is increased in each case. \label{fig:quantumAdvantage}}
\end{figure}

Returning now to the fully quantum case, we would like to compare the performance of quantum single-shot cooling and steady-state refrigeration. Let us define the temperature advantage $\Delta T = T_\infty - \tilde{T}_1(t_0)$ as the difference between the steady-state temperature $T_\infty$ of the cold qubit and the quantum single-shot cooling minimum $\tilde{T}_1(t_0)$. This is an important figure of merit quantifying the advantage gained from quantum single-shot cooling compared to steady-state cooling. Once the cold qubit is decoupled from the refrigerator at time $t_0$, its temperature increases as it equilibrates with the environment. The advantage gained from quantum single-shot cooling lasts only until the temperature of the cold qubit grows larger than the steady-state temperature, which occurs at a time $t_1$. This motivates us to define the quantum advantage time $t_Q = t_1 - t_0$, which is another important quantity characterising the performance of the quantum single-shot refrigerator compared to its steady-state counterpart. See Fig.~\ref{fig:quantumAdvantage}(a) for a graphical depiction of these quantities. 

For simplicity, in this section we work in the strong-coupling limit of Model I, where there always exists some quantum single-shot cooling advantage and the optimal switch-off time is given approximately by $t_0 \approx \pi/(2g)$. In this case, the quantum advantage time $t_Q$ tends to decrease as either of the bath temperatures $T_r$ and $T_h$ is increased. Of course, as the room temperature is varied, the rate of equilibration of the cold qubit with its environment changes, which in turn affects the value of $t_Q$. In order to remove this trivial dependence on $T_r$, we measure the quantum advantage time in units of the relaxation time $\gamma_r^{-1}$, where $\gamma_r=[\gamma_1(E_1) + \gamma_1(-E_1)]/2$ is the arithmetic mean of the thermal gain and decay rates acting on the cold qubit in isolation~\bibnote{Mathematically, $\gamma_r$ is the spectral gap of the Liouvillian $\mathcal{L}_1$ governing the dynamics of the cold qubit after it is decoupled from the refrigerator. That is, $\gamma_r$ is the real part of the eigenvalue(s) of $\mathcal{L}_1$ with the largest (negative) real part, which determines the asymptotic decay rate towards equilibrium.} (the explicit expressions for $\gamma_1(\pm E_1)$ are given in Appendix~\ref{appendixA}).

For a given set of fridge energies $\{E_i,g\}$ and a fixed room temperature $T_r$, one would like to optimise the temperature of the free energy source $T_h$ to obtain the largest fractional temperature advantage $\Delta T/T_\infty$ and also the longest possible advantage time $t_Q$. We show in Fig.~\ref{fig:quantumAdvantage}(b) that there exists a trade-off between these two quantities, which cannot be simultaneously maximised. This trade-off is reminiscent of the competing goals of efficiency and power when optimising thermal machines that operate in the steady-state. Interestingly, there exists a ``sweet spot" where the fractional temperature advantage is maximised, while the advantage time remains adequately large. This immediately suggests a favorable operating regime in which the quantum single-shot cooling protocol is preferable to steady-state cooling. However, in a practical setting the optimum refrigerator parameters will of course depend on the specific application in question. Finally, it is worth noting that introducing coherence in the transport subspace into the initial quantum state further enhances the fractional temperature advantage for a given $t_Q$, thus coherence is also a useful resource in this context.

\section{Conclusions}

We have studied the dynamical evolution of three-qubit absorption refrigerators in different regimes, using two different models of thermal dissipation in open quantum systems. In both these models we encounter oscillations in the temperature of the target qubit below the steady-state temperature, reinforcing the notion of a universal and robust feature of quantum refrigerators. We show that with the right timing these oscillations can be exploited in a quantum single-shot cooling protocol to yield a constant stream of cold qubits.  Numerous quantum information processing protocols require (approximately) pure input states~\cite{NielsenChuang}, potentially making quantum thermal machines a useful addition as pre-cursors to information processing protocols. 

Experimental implementation of quantum single-shot cooling calls for the strong-coupling limit, in which the rate of coherent energy transport exceeds the rate of thermal dissipation. This parameter regime may be difficult to achieve in currently proposed set-ups~\cite{Chen2012epl,Mari2012prl,Venturelli2013prl}. Nevertheless, we hope that our results will serve as an encouragement for further theoretical and experimental work towards achieving strong coupling in quantum absorption refrigerators. Note that a recent work has found evidence that quantum single-shot cooling is even possible in the weak-coupling limit \cite{BohrBrask2015}, providing yet more motivation to experimentally study transient effects in quantum thermal machines. 

The advantage of the absorption refrigerator lies in its high degree of autonomy: no external energy source is required to keep the machine running. Since the initial states we consider come for free in the context of thermodynamic resource theories~\cite{HorodeckiOppenheim2013}, or more explicitly come directly from their respective baths, two thermal reservoirs at a given temperature are sufficient to let the machine run for as long as their temperatures do not change significantly (which for macroscopic baths in contact with quantum systems is sufficiently long for all practical purposes). 

Note that our scheme is not completely autonomous, because we have assumed that the qubit-qubit interaction can be switched on and off by external control. However, it is also possible to envisage a scenario where the switching is performed by a quantum clock~\cite{Malabarba2015njp}. In this case, the global Hamiltonian would be time-independent and the protocol becomes fully autonomous. However, a more detailed study is required in order to understand how a quantum mechanical clock may be used to control the machine: this will form the topic of a future publication. 

Apart from the prospect of information processing, our results help to elucidate the quantum nature of thermodynamics. Compared to its classical counterpart, thermodynamics at the quantum scale often results in additional constraints and limitations~\cite{Horodecki2013a,Brandao2015} due to the discretised nature of the fundamental states. The central question in this context concerns the actual impact of coherence, entanglement and other genuine quantum features on the potential transformations in quantum thermodynamics. Indeed, researchers have only just recently started investigating coherences in the context of resource theories, and a complete understanding is still elusive~\cite{Lostaglio2015}. As we show using the stochastic absorption refrigerator, weakly coupled three-qubit fridges operating in the steady state are in some sense equivalent to classical ones, as one can achieve identical steady-state temperatures from a classical stochastic model. On the other hand, the coherent transport of energy inducing oscillations in the population of the cold qubits constitutes a unique quantum feature. This points towards the potential of harnessing genuine quantum resources to take thermal machines beyond the classically possible. 

\section*{Acknowledgements}
We acknowledge inspiring discussions with Jonatan Bohr-Brask, Nicolas Brunner, Karen Hovhannisyan, Marti Perarnau and Martin Plenio. We also thank Martin Plenio for helpful comments on the manuscript. M.T.M. is grateful to the LIQUID institute for their kind hospitality during the completion of this work, and acknowledges financial support from the UK EPSRC via the Controlled Quantum Dynamics CDT. M.W. acknowledges funding from the Singaporean Ministry of Education, Tier 3 Grant Random numbers from quantum processes (MOE2012-T3-1-009).  J.P. acknowledges funding by the Spanish Ministerio de Econom\'{i}a y Competitividad under Project No. FIS2012-30625. M.H. acknowledges funding from the Juan de la Cierva fellowship (JCI 2012-14155), the European Commission (STREP ``RAQUEL") and the Spanish MINECO Project No. FIS2013-40627-P, the Generalitat de Catalunya CIRIT Project No. 2014 SGR 966. Data underlying work funded by EPSRC can be found in a MATLAB file on the arXiv preprint server (arXiv:1504.01593 [quant-ph]).

\bibliographystyle{unsrt}
\bibliography{SingleShotFridge}
\newpage
\appendix

\section{Thermalisation models}
\label{appendixA}

In this Appendix we describe the two models of thermalisation used throughout this work, and derive the appropriate master equations.

\subsection{Model I}

In this model, each heat bath is represented as a collection of harmonic oscillators, so that the total Hamiltonian for the three baths is $H_B = \sum_{i=1}^3 H_{B_i}$, with
\begin{equation}
\label{bathH}
H_{B_i} =\sum_{\mathbf{k}}\nu_{i,\mathbf{k}}b^{\d}_{i,\mathbf{k}} b_{i,\mathbf{k}},
\end{equation}
where the bosonic mode operators satisfy canonical commutation relations $[b_{i,\mathbf{k}},b^{\d}_{j,\mathbf{k}^{\prime}}] = \delta_{ij}\delta_{\mathbf{k}\mathbf{k}^{\prime}}$ and $[b_{i,\mathbf{k}},b_{j,\mathbf{k}^{\prime}}] = 0$. The qubit-bath interaction is given by
\begin{equation}
\label{qubitBathH}
H_{AB} = \sum_{i=1}^3 A_i \otimes X_i
\end{equation}
where $A_i = \sigma^x_i$ and the collective bath coordinates are defined by
\begin{equation}
\label{bathCoordinate}
X_i = \sum_{\mathbf{k}}\left (\lambda_{i,\mathbf{k}} b_{i,\mathbf{k}} + \lambda_{i,\mathbf{k}}^{\ast}b^{\d}_{i,\mathbf{k}}\right ),
\end{equation}
with constants $\lambda_{i,\mathbf{k}}$ that control the strength of the coupling of qubit $i$ to its associated bath.

\subsubsection{Strong-coupling limit}

We now sketch the derivation of the strong-coupling master equation, valid when $g\gtrsim E_i$. This master equation describes dissipation as resulting from incoherent transitions between the eigenstates of the full coupled Hamiltonian $H_A = H_\mathrm{loc} + V$. When $g=0$, these eigenstates are simply the computational basis states. When $g\neq 0$, the interaction splits the degenerate states spanning the transport subspace into two new eigenstates, denoted by $|\pm\rangle = (|101\rangle \pm |010\rangle)/\sqrt{2}$, with corresponding energy eigenvalues $E_2 \pm g$. The remaining eigenstates and eigenvalues are left unchanged. 

Working in an interaction picture with respect to $H_A + H_B$, the time evolution of the system coupling operators is given by $$A_i(t) = \ee^{\ii H_At} A_i \ee^{-\ii H_A t}.$$ We decompose this into Fourier components as
\begin{equation}
\label{fourierSystemOp}
A_i(t) = \sum_{\omega} \ee^{-\ii\omega t} A_i(\omega), \quad [H_A,A_i(\omega)] = -\omega A_i(\omega),
\end{equation}
where the Bohr frequencies $\{\omega\}$ denote the set of all possible (positive and negative) energy differences between the eigenvalues of $H_A$.

We assume that the initial state of the system factorises as $\rho(0) = \rho_A(0) \bigotimes_{i=1}^3 \rho_{B_i}$, where $\rho_{B_i} =  \ee^{-\beta_i H_{B_i}}/\Tr\left(\ee^{-\beta_i H_{B_i}}\right) $, and that the system-bath coupling is sufficiently weak that perturbation theory can be applied. The master equation is derived by projecting the interaction-picture von Neumann equation for the density operator onto the subspace spanned by states of the form $\rho_A\otimes \rho_{B}$, where $\rho_A = \Tr_B(\rho)$, and truncating the resulting equation at second order in the system-bath coupling (Born approximation). The Markov approximation then consists of assuming that the memory time of the bath is much shorter than all typical time scales of the reduced qubit evolution, see Ref.~\cite{Breuer2007book} for details. Tracing over the baths results in the following equation of motion for the qubit density operator $\tilde{\rho}_A$ in the interaction picture:
\begin{align}
\label{redfieldEquation}
\frac{\dd \tilde{\rho}_A}{\dd t} = \sum_{\omega,\omega^{\prime}} \sum_{i=1}^3 & \ee^{\ii(\omega^{\prime} - \omega)t}\Gamma_i(\omega) \left [ A_i (\omega)\tilde{\rho}_A(t) A^{\d}_i(\omega^{\prime})\right . \notag \\  & \left .- \;A^{\d}_i(\omega^{\prime}) A_i(\omega) \tilde{\rho}_A(t)\right ] + \mathrm{h.c.},
\end{align}
where we defined the self-energy
\begin{equation}
\label{selfEnergy}
\Gamma_{i}(\omega) = \int_{0}^{\infty}\dd t\; \ee^{\ii\omega t} \langle X_i^{\d}(t) X_i(0)\rangle,
\end{equation}
with $X_i(t) = \ee^{\ii H_{B_i} t} X_i \ee^{-\ii H_{B_i}t}$. In general, the self-energy can be written $\Gamma_i(\omega) = \frac{1}{2}\gamma_i(\omega) + \ii S_i(\omega)$ where the real part $\gamma_i(\omega)$ corresponds to an incoherent transition rate, and the imaginary part $S_i(\omega)$ corresponds to an energy shift which we assume to be negligibly small. 

We now perform the rotating-wave approximation by averaging over the oscillating terms in Eq. \eqref{redfieldEquation}, so that terms with $\omega\neq \omega^{\prime}$ drop out. This approximation is valid when the typical energy differences are much larger than the incoherent transition rates, i.e.\ $\min\{E_i,g\} \gg \max\{\gamma_i(\omega)\}$. Transforming back to the Schr\"odinger picture results in a Lindblad equation of the form 
\begin{equation}
\label{RWAmasterEquation}
\frac{\dd \rho_A}{\dd t} = -\ii [H_A,\rho_A] + \sum_{i=1}^3\sum_{\omega}\gamma_i(\omega) \mathcal{D}[A_i(\omega)]\rho_A,
\end{equation}
where the dissipators are given by 
\begin{equation}
\label{dissipatorDef}
\mathcal{D}[L]\rho= L\rho L^{\d} - \frac{1}{2}\{L^{\d}L,\rho\}
\end{equation}
for a general Lindblad operator $L$. Explicitly, the Lindblad operators are
\begin{align*}
A_1(E_1) &= \ket{011}\bra{111} + \ket{000}\bra{100} \\
A_1(E_1+g) & =  \frac{1}{\sqrt{2}}\big (\ket{001}\bra{+} - \ket{-}\bra{110}\big )  \\
A_1(E_1-g) & = \frac{1}{\sqrt{2}}\big (\ket{+}\bra{110} + \ket{001}\bra{-}\big )
\end{align*}
\begin{align*}
A_2(E_2) &= \ket{100}\bra{110} + \ket{001}\bra{011} \\
A_2(E_2+g) & = \frac{1}{\sqrt{2}}\big (\ket{000}\bra{+} + \ket{-}\bra{111}\big ) \\
A_2(E_2-g) & =  \frac{1}{\sqrt{2}}\big ( \ket{+}\bra{111} - \ket{000}\bra{-} \big ) 
\end{align*}
\begin{align*}
A_3(E_3) &= \ket{110}\bra{111} + \ket{000}\bra{001} \\
A_3(E_3+g) & =  \frac{1}{\sqrt{2}}\big (\ket{100}\bra{+} - \ket{-}\bra{011}\big ) \\
A_3(E_3-g) & =  \frac{1}{\sqrt{2}}\big (\ket{+}\bra{011} + \ket{100}\bra{-}\big ),
\end{align*}
while the remaining non-zero Lindblad operators, corresponding to the reverse processes, are found from $A_i(-\omega) = A_i(\omega)^{\dagger}$.

In order to actually evaluate the rates $\gamma_i(\omega)$, we introduce the spectral function of each bath:
\begin{equation}
J_i(\omega) = 2\pi \sum_{\mathbf{k}} \lvert \lambda_{i,\mathbf{k}}\rvert^2 \delta(\omega - \nu_{i,\mathbf{k}}).
\end{equation}
In the limit of an infinite bath with a smooth density of states, the sum over the quantum numbers $\mathbf{k}$ can be approximated by an integral, and $J_i(\omega)$ becomes a continuous function. We assume that the baths have identical spectral functions of the Ohmic form
\begin{equation}
\label{ohmicSpectralFunction}
J(\omega) = \alpha  \omega \ee^{-\omega/\Omega},
\end{equation}
where $\alpha$ is a dimensionless coupling parameter and $\Omega$ is a cut-off frequency of the system-bath interaction, which must be much larger than all other energy scales in order for the Markov approximation to hold. The incoherent rates are then given by 
\begin{equation}
\label{incoherentRates}
\gamma_i(\omega) = \left \lbrace \begin{array}{cc}
J(\omega)[1+n(\omega,\beta_i)]\quad & (\omega>0) \\
J(|\omega|)n(|\omega|,\beta_i) \quad & (\omega<0) 
\end{array}\right. ,
\end{equation}
where $n(\omega,\beta) = (\ee^{\beta \omega} - 1)^{-1}$ denotes the Bose-Einstein distribution. 

\subsubsection{Weak-coupling master equation}

We now consider the weak-coupling limit where $g$ is comparable to the dissipation rates. In this case, the previous derivation is no longer valid since the rotating-wave approximation does not apply to counter-rotating terms of frequency $(\omega-\omega^{\prime}) \sim g$. Instead we should work in an interaction picture generated by $H_\mathrm{loc} + H_B$ and treat the interaction $V$ between the qubits as a small perturbation\cite{Wichterich2007pre,Rivas2010njp}. The time evolution of the system coupling operators is now given by
$$A_i(t) = \ee^{\ii H_\mathrm{loc} t} A_i \ee^{-\ii H_\mathrm{loc} t},$$ 
with the corresponding Fourier decomposition
$$A_i(t) = \sum_{\omega} \ee^{-\ii\omega t} A_i(\omega), \quad [H_\mathrm{loc},A_i(\omega)] = -\omega A_i(\omega),$$
where the frequencies $\{\omega\}$ represent the eigenvalue differences of $H_\mathrm{loc}$ only.

As before, we assume that the initial state of the system factorises as $\rho(0) = \rho_A(0) \otimes  \rho_{B}$. Now we project the interaction-picture von Neumann equation onto states of the form $\rho_A\otimes \rho_{B}$ and truncate the resulting equation at second order in the qubit-bath interaction \textit{and} the qubit-qubit interaction. We then perform the Markov approximation and trace over the bath variables; see Ref.~\cite{Rivas2010njp} for full details of the derivation. The resulting equation of motion is
\begin{align}
\label{localRedfieldEquation}
\frac{\dd \tilde{\rho}_A}{\dd t} = &-\ii [V,\tilde{\rho}_A] \notag\\& + \Bigg (\sum_{\omega,\omega^{\prime}} \sum_{i=1}^3  \ee^{\ii(\omega^{\prime} - \omega)t}\Gamma_i(\omega) \left [ A_i (\omega)\tilde{\rho}_A(t) A^{\d}_i(\omega^{\prime})\right . \notag \\  & \left . - \;A^{\d}_i(\omega^{\prime}) A_i(\omega) \tilde{\rho}_A(t)\right ] + \mathrm{h.c.}\Bigg ),
\end{align}
where the self-energy $\Gamma_i(\omega)$ is defined by Eq. \eqref{selfEnergy}. We then write $\Gamma_i(\omega)\approx \frac{1}{2}\gamma_i(\omega)$, neglecting the imaginary part corresponding to small energy shifts of the qubit energy splittings. The rotating-wave approximation now consists of crossing off counter-rotating terms with $\omega\neq \omega^{\prime}$, all of which have frequencies of order $E_i$. In order to be consistent with our assumptions, we must have $E_i\gg g$. The resulting master equation in the Schr\"odinger picture is
\begin{align}
\label{localMasterEquation}
\frac{\dd \rho_A}{\dd t} = & -\ii [H_A,\rho_A] \notag \\ & + \sum_{i=1}^3\left [\gamma_i(E_i) \mathcal{D}[\sigma_i^{-}]\rho_A + \gamma_i(-E_i) \mathcal{D}[\sigma_i^{+}]\rho_A\right ],
\end{align}
where the rates are given by Eq. \eqref{incoherentRates}.

\subsection{Model II}

In this model, the baths are represented by three additional fictitious qubits described by Pauli operators $\tau^{x,y,z}_i$, with Hamiltonian
\begin{equation}
\label{fictitiousQubitHamiltonian}
H_F = \frac{1}{2}\sum_{i=1}^3 E_i \tau_i^z.
\end{equation}
We have chosen these qubits to have identical energy splittings to their associated physical qubits, in order to avoid renormalising the physical qubit energy splittings. We introduce a Lindblad dissipator for each fictitious qubit corresponding to damping by a perfectly Markovian (delta-correlated in time) reservoir at temperature $T_i$. For simplicity, we assume identical spontaneous emission rates $\gamma$ for the three fictitious qubits. The coupling to the refrigerator is described by the Hamiltonian
\begin{equation}
\label{bathQubitSystemHamiltonian}
H_{AF} = \sum_{i=1}^{3}\eta \gamma( \sigma^+_i\tau^-_i + \sigma^-_i \tau^+_i),
\end{equation}
where $\sigma^\pm_i = \frac{1}{2}(\sigma^x_i \pm \ii \sigma^y_i)$ and $\tau^\pm_i = \frac{1}{2}(\tau^x_i \pm \ii \tau^y_i)$, while $\eta$ is a small dimensionless parameter. The density operator $\rho_{AF}$ of the six-qubit system is therefore described by the master equation
\begin{align}
\label{fictMasterEquation}
\frac{\dd\rho_{AF}}{\dd t} = & -\ii[H_A + H_F + H_{AF}, \rho_{AF}] \notag \\ & + \gamma \sum_{i=1}^3\left [ \mathcal{D}[\tau_i^{-}]\rho_{AF} + \ee^{-\beta_i E_i} \mathcal{D}[\tau_i^{+}]\rho_{AF}\right].
\end{align}
The effective spectral density seen by each physical qubit is of the Lorentzian form
\begin{equation}
\label{effSpectralDensityFict}
J_i(\omega) = \frac{\eta^2\gamma^2\Gamma_i}{\Gamma_i^2 + (E_i - \omega)^2},
\end{equation}
where $\Gamma_i = \gamma(1+\ee^{-\beta_i E_i})/2$. For a fixed $\eta$ and $T_i$, the bandwidth of the effective bath's frequency response is therefore controlled by modifying the parameter $\gamma$.

\section{Phase sensitivity analysis}
\label{appendixB}

In this Appendix we consider the effect on cooling of coherence in the initial quantum state with a small amount of phase noise. In order to gain some analytical insight, we suppose that the dynamics can be treated as approximately unitary over the time scales of interest. This is a decent approximation in the strong-coupling limit, where $g$ is much larger than the relaxation rate. 

We define the populations $a(t) = \bra{010}\rho_A(t)\ket{010}$ and $b(t) = \bra{101}\rho_A(t)\ket{101}$, and the coherence $\mathcal{C}(t) = \bra{101}\rho_A(t)\ket{010}$, where $\rho_A(t)$ denotes the reduced quantum state of the refrigerator qubits. We also introduce the convenient parametrisations $S = [a(0) + b(0)]/2$, $D = [b(0) - a(0)]/2$ and $\mathcal{C}(0) = \ii r\ee^{\ii\phi}$, where $r>0$ is the magnitude and $\phi$ is the phase. The refrigeration condition $T_v < T_1$, where $T_v$ is the virtual temperature \eqref{effTemp} and $T_1$ is the initial temperature of the cold qubit, implies that $D > 0$. 

At short times, the effect of dissipation can be approximately neglected, and the evolution of the populations $a(t), b(t)$ is analogous to that of a resonantly driven two-level system. In this approximation, the populations at time $t$ are given by
\begin{equation}
 \label{populationDynamics}
 a(t) = S - D \cos(2gt) + r\cos(\phi) \sin(2gt),
 \end{equation} 
and $b(t) = 2S - a(t)$. The population difference of the cold qubit is given by
\begin{equation}
\label{coldQubitPopDiff}
\langle \sigma^z_1 (t) \rangle = \langle\sigma^z_1(0)\rangle  + 2\left [ a(0) - a(t)\right ],
\end{equation}
which is related to the effective temperature by Eq.~\eqref{effTemp}. From this we find that the first temperature minimum occurs at time
\begin{equation}
\label{topt}
t_\mathrm{min}(\phi) = \frac{1}{2g}\left ( \pi - \tan^{-1} \left [\frac{r\cos (\phi)}{D}\right ]\right ).
\end{equation}
with
\begin{equation}
\label{amplitudeT0}
a(t_\mathrm{min}) = S + \sqrt{D^2 + r^2\cos^2(\phi)}.
\end{equation}
The amplitude of the population oscillations is therefore maximised by choosing $\phi=0$, i.e. a purely imaginary initial coherence, as expected from Eq.~\eqref{energyBalance}. With the choice of $\phi = 0$, the optimal time to extract the cold qubit is given by $t_\mathrm{opt} = t_\mathrm{min}(0)$.

We now suppose that each qubit entering the refrigerator is prepared with a random phase $\phi$ of the coherence in the transport subspace, due to experimental noise, for example. These phase shifts are described by some probability distribution function $p(\phi)$. In order to find the dynamics of the resulting ensemble of cold qubits, we average Eq.~\eqref{populationDynamics} over the distribution $p(\phi)$. This procedure leads to the replacement of $\cos(\phi)$ in all expressions by
\begin{equation}
\label{avCosPhi}
\overline{\cos(\phi)} = \int_{-\pi}^{\pi}\dd\phi\, p(\phi) \cos(\phi).
\end{equation}

As a simple but broadly relevant example, we assume that phase shift has zero mean $\overline{\phi} = 0$, and that its variance $\overline{\phi^2}$ is smaller than unity. We also suppose that its cumulants are rapidly decreasing, so that $p(\phi)$ may be approximated by a Gaussian distribution. Under these assumptions, we obtain
\begin{equation}
\label{avCosPhiGaussian}
\overline{\cos(\phi)} = \ee^{-\overline{\phi^2}/2}.
\end{equation}
 We find from Eq.~\eqref{topt} that the minimum temperature of the ensemble occurs at time
\begin{equation}
\label{toptPrime}
t_\mathrm{opt}' = \frac{1}{2g}\left ( \pi - \tan^{-1} \left [\frac{r\ee^{-\overline{\phi^2}/2}}{D}\right ]\right ),
\end{equation}
so that $t_\mathrm{opt}' > t_\mathrm{opt}$ in this approximation. In the following, primed variables indicate quantities in the presence of noise, while the unprimed variables denote the corresponding quantities in the absence of noise.

In order to find the shift in the ensemble temperature due to the phase noise, we consider two distinct scenarios. In the first scenario, we assume that some information about the phase distribution is known, so that it is possible to predict in advance the optimal time $t_\mathrm{opt}'$ to extract the qubit. This information could be obtained by expending some qubits in order to characterise the noisy preparation by quantum state tomography. In this case, the phase noise leads to a change in the population difference of the cold qubit given by
\begin{align}
\label{popDiffWithInfo}
\Delta\langle\sigma_1^z\rangle & = \langle\sigma_1^z(t_\mathrm{opt}')\rangle' - \langle\sigma_1^z(t_\mathrm{opt})\rangle \notag \\
& = 2\sqrt{D^2 + r^2} -  2\sqrt{D^2 + r^2\ee^{-\overline{\phi^2}}} \notag \\
& \approx \frac{r^2}{\sqrt{D^2 + r^2}}\left (1-\ee^{-\overline{\phi^2}}\right ).
\end{align}

In the second scenario, we assume that no information about the phase noise distribution is known, so that the experimenter extracts the cold qubit at the incorrect time $t_\mathrm{opt}$. For example, this scenario applies if only a single qubit is available for cooling, so that the noisy preparation cannot be characterised in advance. In this case, phase noise changes the population difference by the amount
\begin{align}
\label{popDiffNoInfo}
\Delta\langle\sigma_1^z\rangle & = \langle\sigma_1^z(t_\mathrm{opt})\rangle' - \langle\sigma_1^z(t_\mathrm{opt})\rangle \notag \\
& = \frac{2r^2}{\sqrt{D^2 + r^2}}\left (1-\ee^{-\overline{\phi^2}/2}\right ).
\end{align}
Note that it is also possible to consider the case where only partial information about the noise distribution is known, in which case we expect $\Delta\langle\sigma_1^z\rangle$ to lie somewhere between Eqs.~\eqref{popDiffWithInfo} and \eqref{popDiffNoInfo}.

In any case, assuming that the effect of phase noise is small, we find that the expected shift in the ensemble temperature to lowest order in $\Delta\langle\sigma_1^z\rangle$ is
\begin{equation}
\label{TshiftAppendix}
 \Delta \tilde{T}_1 = \frac{2\tilde{T}_1^2}{E_1} \cosh^2\left (\frac{E_1}{2\tilde{T}_1}\right ) \Delta\langle\sigma^z_1\rangle,
\end{equation}
where $\tilde{T}_1$ denotes the optimal temperature in the absence of phase noise. This equation indicates that the effect of phase fluctuations is minimised when $\tilde{T}_1\sim E_1$, while the most drastic effects occur for very large or very small target temperatures. In particular, the right-hand side of Eq.~\eqref{TshiftAppendix} diverges as $\tilde{T}_1 \to 0$, which indicates that even vanishingly small fluctuations in the initial coherence phase may prevent one from using this coherence to cool the cold qubit to its ground state.

\begin{figure}
\includegraphics[scale=0.24]{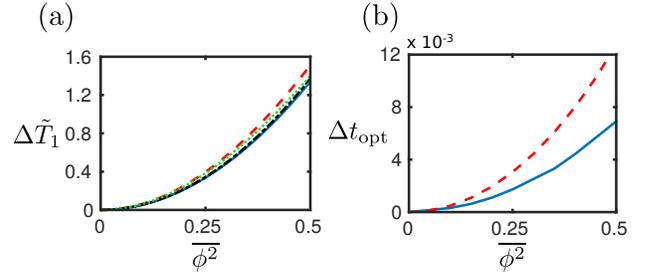}
\caption{Comparison of exact numerical and approximate analytical results for the effect of Gaussian zero-mean phase noise characterised by its variance $\overline{\phi^2}$. The same parameters are used as in Fig.~\ref{Figure2}(a), with 5\% of the maximum initial coherence. (a) Shift in the cold qubit temperature when extracted: numerical (blue solid line) and analytical (red dashed line) results when the phase noise distribution is not known, numerical (black dot-dashed) and analytical (green dotted line) results when the phase noise distribution is known. (b) Shift in the time of the temperature minimum $\Delta t_\mathrm{opt} = t_{\mathrm{opt}}' - t_\mathrm{opt}$: numerical (blue solid line) and analytical (red dashed line) results.\label{gaussianPhaseNoise}}
\end{figure}

In Fig.~\ref{gaussianPhaseNoise}(a) we numerically calculate and plot an example of the change in the final temperature of the cold qubit when it is disconnected from the machine, either with or without sufficient knowledge of the phase noise distribution to predict the optimal extraction time. We see that in fact it makes little difference to the temperature whether or not the optimal extraction time is known precisely, since the shift in this time due to phase noise is rather small (Fig.~\ref{gaussianPhaseNoise}(b)). Our approximate analytical results are also shown in Fig.~\ref{gaussianPhaseNoise} for comparison. 

\section{Stochastic refrigerator model}
\label{appendixC}

In this Appendix we derive an effective stochastic model of the quantum absorption refrigerator in the deep weak-coupling limit. Our analysis follows Ref.~\citep{Kamiya2015ptep}, where it was shown that coherent propagation of a quantum particle on a lattice under strong local dephasing can be approximated by a stochastic hopping process. In our case, the dephasing is provided directly by the action of the thermal baths, which destroy the coherences in the computational basis at a characteristic rate $\Gamma$. In the limit where $g\ll \Gamma$, we can derive a closed equation of motion for the populations, describing their dynamics over times coarse-grained on the scale $\Gamma^{-1}$.
 
Working in the weak-coupling limit of Model I, we begin with the master equation \eqref{localMasterEquation}. This may be written as
\begin{equation}
 \label{weakCouplingMasterEquation}
 \frac{\dd \rho}{\dd t} = \mathcal{L}\rho,
 \end{equation} 
where we introduced the Liouvillian superoperator $\mathcal{L}$, and for brevity we have written $\rho = \rho_A$ for the reduced state of the refrigerator qubits. We decompose the Liouvillian as $\mathcal{L} = \mathcal{L}_0 + \mathcal{V}$, where $\mathcal{V}\rho = \ii[\rho,V]$, while the superoperator $\mathcal{L}_0$ contains both the local Hamiltonian and dissipative contributions to the Liouvillian. 

The quantum state can be expanded in the eigenbasis of $\mathcal{L}$ as
\begin{equation}
\label{rhoExpand}
\rho(t) = \sum_{\lambda} \rho_\lambda \ee^{\lambda t},
\end{equation}
where $\rho_\lambda$ is an eigenvector of $\mathcal{L}$ with complex eigenvalue $\lambda$. Since $\Re(\lambda) \leq 0$, we see that only the eigenvalues with the largest real part are relevant as $t\to \infty$. In this basis the master equation reduces to the eigenvalue equation
\begin{equation}
\label{eigenvalueEqn}
\left (\mathcal{L}_0 + \mathcal{V}\right ) \rho_\lambda	= \lambda	\rho_\lambda.
\end{equation}
The eigenvalues of $\mathcal{L}_0$ are of order $\Gamma$, where $\Gamma$ is a characteristic dephasing rate of the thermal dissipation. On the other hand, the eigenvalues of $\mathcal{V}$ are of order $g$, which is much smaller than $\Gamma$ by assumption. We can therefore treat the effect of $\mathcal{V}$ as a small perturbation.

We introduce a projector $\mathcal{P}$ defined by 
\begin{equation}
\label{populationProjector}
\mathcal{P}\rho = \sum_{n=0}^7 \bra{n}\rho\ket{n} \ket{n}\bra{n},
\end{equation}
where $\{\ket{n}\}$ are the computational basis states. This projects onto the space of populations (diagonal matrix elements) in the computational basis. We also define its orthogonal complement by $\mathcal{Q} = 1 - \mathcal{P}$, which projects onto the space of coherences (off-diagonal matrix elements). We refer to the spaces of populations and coherences as the $\mathcal{P}$-space and the $\mathcal{Q}$-space, respectively. It is readily verified that the following properties hold:
\begin{equation}
\label{PVP}
\mathcal{PVP} = 0,
\end{equation}
\begin{equation}
\label{PLcommutator}
[\mathcal{P},\mathcal{L}_0] = [\mathcal{Q},\mathcal{L}_0] = 0.
\end{equation}
Eq.~\eqref{PVP} states that the interaction $\mathcal{V}$ only couples populations to coherences, while Eq.~\eqref{PLcommutator} reflects the fact that the local Liouvillian $\mathcal{L}_0$ does not couple populations and coherences, i.e.\ it is block-diagonal.

By introducing the identity $1 = \mathcal{Q} + \mathcal{P}$ into both sides of the eigenvalue equation~\eqref{eigenvalueEqn}, we obtain
\begin{align}
\label{Peigenvalue}
\lambda \mathcal{P}\rho_\lambda &= \mathcal{L}_0 \mathcal{P}\rho_\lambda + \mathcal{P}\mathcal{V}\mathcal{Q}\rho_\lambda \\
\label{Qeigenvalue}
\lambda	\mathcal{Q}\rho_\lambda &= \mathcal{L}_0 \mathcal{Q}\rho_\lambda + \mathcal{QVQ}\rho_\lambda + \mathcal{VP}\rho_\lambda,
\end{align}
where Eqs.~\eqref{PVP} and \eqref{PLcommutator} have been used. We now solve Eq.~\eqref{Qeigenvalue} for $\mathcal{Q}\rho_\lambda$ and substitute the result back into Eq.~\eqref{Peigenvalue}, finding
\begin{equation}
\label{Psolution}
\lambda \mathcal{P}\rho_\lambda = \mathcal{L}_0 \mathcal{P}\rho_\lambda + \mathcal{PV}\left ( \lambda - \mathcal{QV} - \mathcal{L}_0 \right )^{-1} \mathcal{VP}\rho_\lambda.
\end{equation}

Since we are looking for eigenvalues $\lambda$ which are small in magnitude, we may neglect the term $\lambda - \mathcal{QV}$, which is of order $g$, in comparison to $\mathcal{L}_0$, which is of order $\Gamma$. Hence we obtain the approximate eigenvalue equation 
\begin{equation}
\label{effLiouvillian}
\lambda \mathcal{P}\rho_\lambda =  \mathcal{L}_\mathrm{eff} \mathcal{P}\rho_\lambda,
\end{equation}
where
\begin{equation}
\label{Leff}
\mathcal{L}_\mathrm{eff} = \mathcal{L}_0 -\mathcal{PVL}_0^{-1}\mathcal{VP}.
\end{equation}
As $t\to \infty$, the master equation \eqref{weakCouplingMasterEquation} can therefore be approximated in the $\mathcal{P}$-space by 
\begin{equation}
\label{LeffME}
 \frac{\dd \rho}{\dd t} = \mathcal{L}_\mathrm{eff}\rho.
\end{equation}

Note that $\mathcal{L}_\mathrm{eff}$ is well defined by Eq.~\eqref{Leff} despite the fact that the unperturbed Liouvillian possesses a zero eigenvector (the steady-state solution): $\mathcal{L}_0\rho_\infty = 0$. This is because the unique steady-state solution lies in the $\mathcal{P}$-space, i.e.\ $\rho_\infty = \mathcal{P}\rho_\infty$. Furthermore, one can write $\mathcal{VP} = \mathcal{QVP}$ by virtue of Eq.~\eqref{PVP}. Therefore, the operator $\mathcal{L}_0^{-1}$ acts only in the $\mathcal{Q}$-space, where all eigenvalues of $\mathcal{L}_0$ are non-zero. 

In order to calculate $\mathcal{L}_\mathrm{eff}$ explicitly, we note that $\mathcal{V}$ annihilates all populations apart from $\ket{010}\bra{010}$ and $\ket{101}\bra{101}$. The action of $\mathcal{L}_\mathrm{eff}$ can therefore be found from its action on these two populations. We also make use of the fact that 
\begin{align}
\label{L1eigenvector}
\mathcal{L}_0\ket{010}\bra{101} & = -\Gamma \ket{010}\bra{101} \notag \\ 
\mathcal{L}_0\ket{101}\bra{010} & = -\Gamma \ket{101}\bra{010},
\end{align}
with 
\begin{equation}
\label{bigGamma}
\Gamma = \frac{1}{2}\sum_{i=1}^3\left [\gamma_i(E_i) + \gamma_i(-E_i)\right ],
\end{equation}
where $\gamma_i(E_i)$ is defined by Eq.~\eqref{incoherentRates}. Using these properties, a straightforward calculation yields the result 
\begin{equation}
\label{L1explicit}
\mathcal{L}_\mathrm{eff}= \mathcal{L}_0 + \frac{2g^2}{\Gamma} \mathcal{D}[B] + \frac{2g^2}{\Gamma} \mathcal{D}[B^\d],
\end{equation}
where $B = \ket{010}\bra{101}$, and $\mathcal{D}$ is defined by Eq.~\eqref{dissipatorDef}. This effective Liouvillian describes transport as a stochastic process, which transfers population symmetrically between the states $\ket{010}$ and $\ket{101}$ at a rate $2g^2/\Gamma$.
\end{document}